\documentclass[aps,twocolumn,showpacs]{revtex4}
\usepackage{graphicx}
\usepackage{amsmath}
\usepackage{amsfonts}
\usepackage{color}

\begin{document}

\title{Stability mapping of bipartite tight-binding graphs with losses and gain: ${\cal PT}-$symmetry and beyond} 

\author{L. A. Moreno-Rodriguez,$^1$ C. T. Martinez-Martinez,$^2$ J. A. Mendez-Bermudez,$^1$ and Henri Benisty$^{3,4}$}

\address{
$^1$Instituto de F\'{\i}sica, Benem\'erita Universidad Aut\'onoma de Puebla,
Apartado Postal J-48, Puebla 72570, Mexico \\
$^2$Universidad Aut\'onoma de Guerrero, Centro Acapulco CP 39610,
Acapulco de Ju\'arez, Guerrero, Mexico\\
$^3$Laboratoire Charles Fabry, Institut d'Optique Graduate School, CNRS, Univ. Paris Saclay, 2 Av. Augustin
Fresnel, 91127 Palaiseau Cedex, France \\
$^4$Universit\'e de Paris, LIED, CNRS UMR 8236, 5 Rue Thomas Mann, 75013 Paris, France
}

\date{\today} \widetext

\begin{abstract}
We consider bipartite tight-binding graphs composed by $N$ nodes split into two sets of equal size: 
one set containing nodes with on-site loss, the other set having nodes with on-site gain.
The nodes are connected randomly with probability $p$.
We give a rationale for the relevance of such ``throttle/brake'' coupled systems (physically open systems) to grasp the stability issues of complex networks in areas such as biochemistry, neurons or economy, for which their modelling in terms of non-hermitian Hamiltonians is still in infancy. Specifically, 
we measure the connectivity between the two sets with the parameter $\alpha$, which is the ratio of 
current adjacent pairs over the total number of possible adjacent pairs between the sets.
For general undirected-graph setups, the non-hermitian Hamiltonian $H(\gamma,\alpha,N)$ of this model 
presents pseudo-Hermiticity, where $\gamma$ is the loss/gain strength.
However, we show that for a given graph setup $H(\gamma,\alpha,N)$ becomes ${\cal PT}-$symmetric.
In both scenarios (pseudo-Hermiticity and ${\cal PT}-$symmetric), depending on the parameter 
combination, the spectra of $H(\gamma,\alpha,N)$ can be real even when it is non-hermitian.
Thus, we numerically characterize the average fractions of real and imaginary eigenvalues of $H(\gamma,\alpha,N)$ 
as a function of the parameter set $\{\gamma,\alpha,N\}$.
We demonstrate, for both setups, that there is a well defined sector of the $\gamma\alpha-$plane (which grows with $N$) where the spectrum of $H(\gamma,\alpha,N)$ is predominantly real.
\end{abstract}

\maketitle

\section{Introduction}

Non-hermitian Hamiltonians (NHH) have become one of the privileged windows to grasp the physics of open systems and their non-equilibrium dynamics. Apart from the actual thermodynamic aspects, a salient issue of NHH is whether these operators possess real or imaginary eigenvalues. Gain and loss are the simplest physical depictions associated with non-hermitian contribution. A logical setting consists in having associated imaginary entries on the diagonal of the operator. The Hamiltonian then intends to describe coupled systems with real coupling, but with each unit (related to each vector component of the system) being liable to operate with gain or loss, with an associated energy flow directed inward from, or outward to, a reservoir, and with an explicit and essentially deterministic amplitude coupling to other relevant units.

In various areas of complex systems (biology, ecology, economy, etc., see Refs.~\cite{Aldridge2006b,Cilib2007,Vanb2009,Murphy2009,Hennequin2014} for instance), energy producing units and energy consuming units can be reasonably well defined, and interact together. As pioneered by R. M. May in 1972~\cite{May1972}, an ``agnostic" description of the many interactions of complex systems can be appropriately tackled through random matrix theory (RMT), a powerful tool with vast application fields nowadays (e.g.~in economy~\cite{Allesina2015}). 
The overall evolution of a complex system can be very different according to the real or complex nature of the corresponding eigenvalues. Example of ``stabilization", steered by the collapse of eigenvalues on the appropriate axis, have been made explicit e.g.~for real neural networks~\cite{Hennequin2014}. The issue also transpires in the topic of parity-time symmetry (notably in optics), that blossomed in the previous decade. While the basic setting of a single gain and a single loss unit has been a staple of pioneer studies, the shift to multi-unit systems is increasingly discussed, with possible relation to ``lasers on graphs"~\cite{Gaio2019,Rotter2019} and recently discussed photonic systems that are liable to rest on coupled units partly endowed with gain~\cite{Lin2020,Beni2015}. Indeed, photonics, e.g. in integrated form, lends itself very well to the setup of canonical complex systems with gain and loss, easily controlled by known electro-optical devices.

Given the overall incomplete knowledge on the field, it is advisable to make some simplifying assumptions. E.g., in Ref.~\cite{MMMB19,BG20}, the coupling between gain or between loss units was deemed possible, which made evolution and regime transitions rather fuzzy. There are various reasons to assume that bipartite systems, whereby gain and loss units are not coupled among themselves but are freely coupled to each other, are a useful basis, inducing the emergence of more clear-cut phenomena. As can be guessed from the abovementioned papers on signalling pathways inside cells, in biology, the very shape and stereochemistry of interaction is chosen to privilege various couplings, with myriads of enzymes that channel processes and separate in- and out-flow of energy and material. Economy offers also a traditional ``bipartite'' view of consumers and producers, especially at both ends of the spectrum (primary energy to useful energy converters: producers; and at the other end, pure consumers paid for their workforce). We anticipate that econophysics could profit from insights from gain-loss coupled systems and their stability, a recurrently debated issue since a century (by Keynes, Minsky, and many others), as the real fate of economy casts doubts on the degree of relevance of ``general equilibrium" approaches.

The bipartite graph situation thus demands more understanding, in terms of addressing the real or imaginary eigenvalues distribution that could be meaningful for exemplary settings in several disciplines.

Here we address the issue of the fraction of each kind of eigenvalues (real and imaginary) as a function of the parameters that logically emerge from the joint consideration of RMT and bipartite (undirected) tight-binding random graphs represented by NHH. On the one hand, for a canonical ordering of the partition, the graphs are represented by off-diagonal (randomly-weighted) block adjacency matrices and are also classically parametrized by a sparsity parameter. On the other hand, the units have gain or loss, represented by imaginary contributions on the diagonal entries of the Hamiltonian matrix, one half with a gain sign, the other half with loss.

Then, the issue that must be addressed is what classes of NHH produce eigenvalues that can stay real. After a preliminary section (Sec.~\ref{pre}) describing the formalism, we define in  Sec.~\ref{properties} the two classes of NHH we shall consider: ${\cal PT}-$symmetric ones and pseudo-Hermitian ones, that correspond to different relationship of the two off-diagonal blocks (either identical or transposed) of the graph adjacency matrix.
We then solve the four-unit problem, the sole to be analytically tractable.
By numerical means, in Sec.~\ref{spectra}, we then describe the statistics of the average fractions of real and imaginary eigenvalues of NHH as a function of the model parameters.
In particular we show, for both ${\cal PT}-$symmetric and pseudo-Hermitian Hamiltonians, that there is a well defined sector in the parameter space (which grows with the graph size) where the spectrum is predominantly real, thus favoring stability.
In the discussion part of the paper, Sec.~\ref{conclusions}, we summarize the main findings and give a hint of the case of a cousin binary system (where the off-diagonal elements of the NHH are 1 or 0 as in standard adjacency matrices). It also evokes a string of less-common photonic studies whereby remarkable dispersion properties of periodic waveguides come into play~\cite{Ostrovsky1995,Ostrovsky1995b,BenistyPRA2011,KurtJOSA2008,KhayamPNFA2010}.

\section{Preliminaries}
\label{pre}

We consider undirected bipartite tight-binding random graphs $G$ composed by $N$ nodes split into two sets of equal size: 
the set $V_{+}(G)$ contains nodes with on-site loss and the set $V_{-}(G)$ has nodes with on-site gain. 
The nodes are connected randomly with probability $p$. We characterize the connectivity of the graph 
with the parameter $\alpha$ which is the ratio of current adjacent pairs over the total number of possible 
adjacent pairs between the sets, thus $\alpha\in[0,1]$.
The tight-binding Hamiltonian of our bipartite graphs can be written as  
\begin{eqnarray}
\label{TBH}
H(\gamma,\alpha,N) = i\gamma \sum_{n\in V_{+}(G)} | n \rangle \langle n| -
 i\gamma \sum_{n\in V_{-}(G)} | n \rangle \langle n| \nonumber \\ +
\sum_{n\in V_{+}(G)} \ \sum_{m\in V_{-}(G)} \left( u_{nm} | n \rangle \langle m| + u_{mn} |m \rangle \langle n| \right) \ ,
\end{eqnarray}
where $\gamma\in \mathbb{R}^+$ is the loss/gain strength and $u_{mn}$ are the hopping 
integrals between sites $n$ and $m$. 
Also, $u_{nm}=u_{mn}$, since the graph is assumed as undirected. 
As for standard RMT ensembles, we choose $u_{nm}$ as statistically 
independent random variables drawn from a normal distribution with zero mean and unit variance.

Equation~(\ref{TBH}) can also be conveniently written as
\begin{eqnarray}
H(\gamma,\alpha,N) = H_\gamma(\gamma,N) + H_\alpha(\alpha,N) ,
\label{eq:1}
\end{eqnarray}
with
$$
   H_\gamma(\gamma,N) = \;
    i \gamma \begin{pmatrix}
     I_{N/2} & 0  \\ 
     0 & - I_{N/2}
   \end{pmatrix}_{N\times N} 
$$
and
$$
H_\alpha(\alpha,N) = \;
   \begin{pmatrix}
      0 & U  \\ 
     U^T & 0
   \end{pmatrix}_{N\times N}  .
$$
Here $U$ and $U^T$ are real $N/2 \times N/2$ matrices which depend on the parameter $\alpha$:
for $\alpha=0$ both are null matrices, 
when $\alpha=1$ both are full random matrices,
while for $\alpha\in (0,1)$ they are sparse random matrices.

Note that $H_\alpha(\alpha,N)$ is in fact the randomly-weighted adjacency matrix of the binary network. Also note that $H_\alpha(\alpha,N)$ is similar to the randomly-weighted adjacency matrix {\it with loops} used to study the spectral properties of binary networks in~\cite{MAMPS19}.

\begin{figure*}
\begin{center} 
\includegraphics[width=0.7\textwidth]{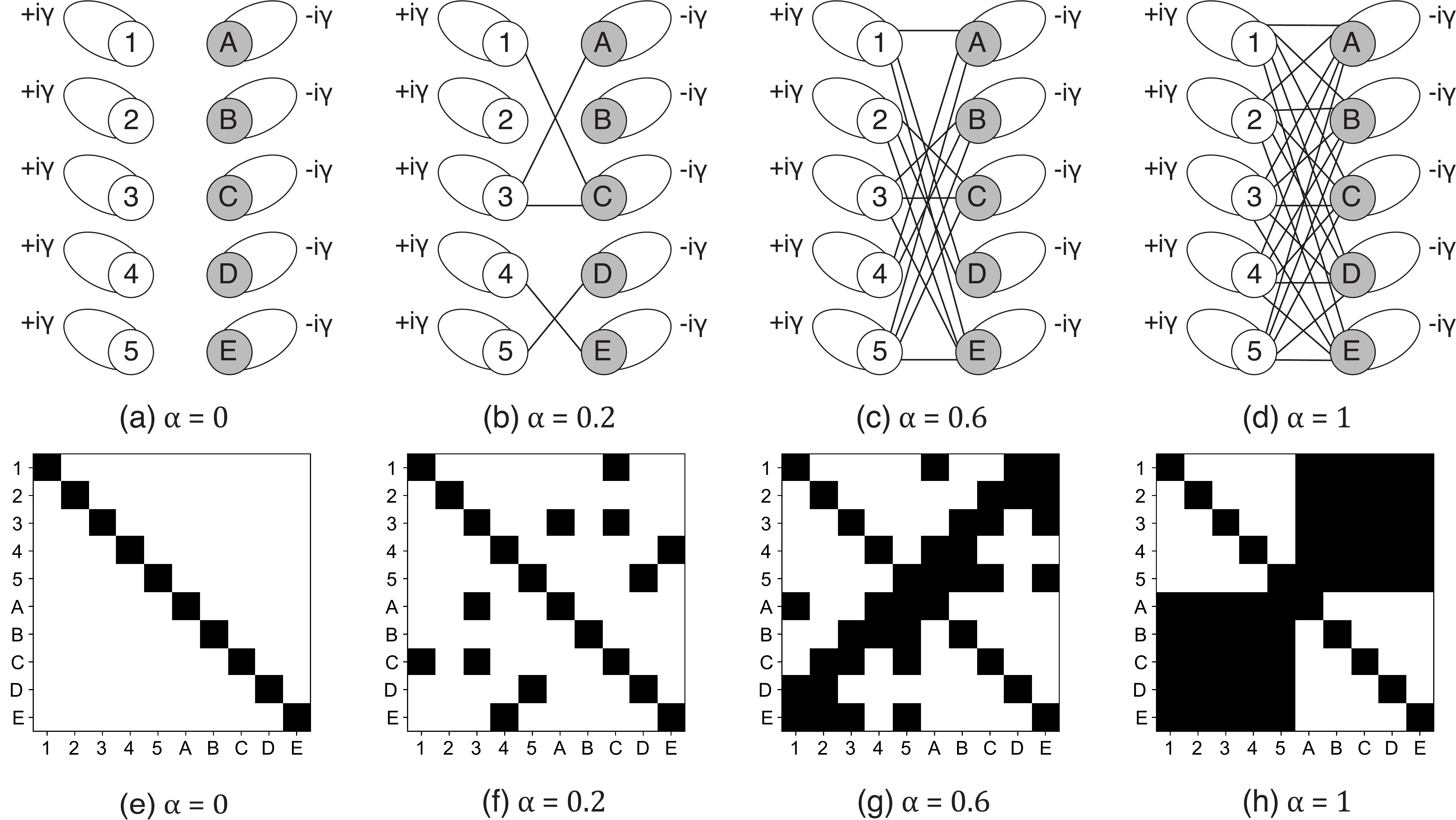}
\caption{(a-d) Examples of ${\cal PT}-$symmetric bipartite graphs. 
(e-h) Non-zero elements of the adjacency matrices corresponding to the graphs in the upper panels.
Here we choose $N=10$ with (a,e) $\alpha=0$, (b,f) $\alpha=0.2$, (c,g) $\alpha=0.6$, and (d,h) $\alpha=1$.}
\label{Fig01}
\end{center}
\end{figure*}

\section{Properties of $H(\gamma,\alpha,N)$}
\label{properties}

It is important to stress that even when the Hamiltonian of Eq.~(\ref{eq:1}) represents a graph with balanced
loss and gain, it is in general non ${\cal PT}-$symmetric.
However, by the proper choice of the graph setup the ${\cal PT}-$symmetry can indeed be imposed to 
$H(\gamma,\alpha,N)$, as it is shown below.

Recall that a Hamiltonian $H$ is ${\cal PT}-$symmetric if the relation $[{\cal PT}, H] = 0$ is satisfied.
Here, ${\cal P}$ is the parity operator, which in matrix form is the first Pauli matrix $\sigma_x$, and ${\cal T}$ 
is the time-reversal operator or the complex conjugation operator.
Thus, for our tight-binding Hamiltonian we have
\begin{eqnarray}
{\cal PT} H - H {\cal PT} & = & 
   \begin{pmatrix}
      0 & 1  \\ 
      1 & 0
   \end{pmatrix}   
      \begin{pmatrix}
     i \gamma I_{N/2} & U \\
     U^T & - i \gamma I_{N/2}
   \end{pmatrix}^{*} \nonumber \\ & & -
    \begin{pmatrix}
      i \gamma I_{N/2} & U  \\ 
      U^T & -  i \gamma I_{N/2}
   \end{pmatrix} 
         \begin{pmatrix}
     0 & 1  \\ 
     1 & 0
   \end{pmatrix} 
	\nonumber \\ & = &
   \begin{pmatrix}
      U^T - U & 0 \\
      0 &  U - U^T 
   \end{pmatrix} .
\label{eq4}
\end{eqnarray}
Since in general $U \neq U^T$, $[{\cal PT}, H] \neq 0$ and $H(\gamma,\alpha,N)$ is not 
${\cal PT}-$symmetric.

\subsection{${\cal PT}-$symmetric Hamiltonian}

According to \eqref{eq4}, the Hamiltonian $H(\gamma,\alpha,N)$ becomes ${\cal PT}-$symmetric when $U=U^{T}$,
which implies that the matrix $U$ is symmetric. In Fig.~\ref{Fig01} we show some examples of ${\cal PT}-$symmetric 
bipartite tight-binding graphs together with their corresponding adjacency matrices; in this figure the trivial cases 
$\alpha=0$ and $\alpha=1$ are also included as a reference.
In this case, we expect to observe regions in the parameter space of our bipartite graph model where the 
spectrum of $H(\gamma,\alpha,N)$ is real even when it is non-hermitian.
We will numerically explore some spectral properties of $H(\gamma,\alpha,N)$ in Subsection~\ref{SS_PT}.

Moreover, in the ${\cal PT}-$symmetric case, $H(\gamma,\alpha,N)$ displays three symmetries~\citep{SHEK12}:
\begin{itemize}
\item[(i)] time-reversal symmetry: 
$$
    H^{T} =
   \begin{pmatrix}
      i \gamma I_{N/2} &  U  \\ 
      U & - i \gamma I_{N/2}
   \end{pmatrix}^{T} 
   = 
      \begin{pmatrix}
     i \gamma I_{N/2} & U  \\ 
     U &  - i \gamma I_{N/2}
   \end{pmatrix}  = H,
$$

\item[(ii)] chiral symmetry:  
\begin{eqnarray*}
\sigma_{z} H \sigma_{z} & = &
   \begin{pmatrix}
      1 & 0  \\ 
      0 & -1
   \end{pmatrix}   
      \begin{pmatrix}
     i \gamma I_{N/2} & U  \\ 
     U & - i \gamma I_{N/2}
   \end{pmatrix}
    \begin{pmatrix}
      1 & 0  \\ 
      0 & -1
   \end{pmatrix}  \\ & = & - \;
   \begin{pmatrix}
      - i \gamma I_{N/2} & U \\ 
      U & i \gamma I_{N/2}
   \end{pmatrix} = - H^{\dagger}
\end{eqnarray*} 
and

\item[(iii)] particle-hole symmetry: 
\begin{eqnarray*}
\sigma_{z} H^{*} \sigma_{z} & = &
   \begin{pmatrix}
      1 & 0  \\ 
      0 & -1
   \end{pmatrix}   
      \begin{pmatrix}
   -  i \gamma I_{N/2} & U  \\ 
     U &  i \gamma I_{N/2}
   \end{pmatrix}
    \begin{pmatrix}
      1 & 0  \\ 
      0 & -1
   \end{pmatrix}  \\ & = &  - 
   \begin{pmatrix}
       i \gamma I_{N/2} & U \\ 
      U & - i \gamma I_{N/2}
   \end{pmatrix}   = - H \, .
\end{eqnarray*} 
\end{itemize}
Particle-hole symmetry (or, equivalently, chiral and time-reversal symmetry) implies that the complex 
eigenvalues of $H(\gamma,\alpha,N)$ come in pairs~\citep{M20}, i.e.~$(\lambda,-\lambda^*)$, which are symmetric 
with respect to the imaginary axis.

\subsection{Pseudo-Hermitian Hamiltonian}

For general undirected graphs, $U \neq U^{T}$ and $H(\gamma,\alpha,N)$ is not ${\cal PT}-$symmetric, 
however, it shows pseudo-Hermiticity:
A matrix has pseudo-Hermiticity symmetry if it satisfies the condition $H = q H^{\dagger} q^{-1}$, where 
$q$ is a Hermitian operator, referred to as the \textit{metric operator}, such that $q^{\dagger} q^{-1}=1$.  
Indeed, the non ${\cal PT}-$symmetric Hamiltonian $H(\gamma,\alpha,N)$ is pseudo-hermitian with respect 
to the first Pauli matrix $q\equiv \sigma_{x}$ or, in other words, $H(\gamma,\alpha,N)$ is pseudo-Hermitian with 
respect to the parity operator $P$:
\begin{eqnarray*}
\sigma_{x} H \sigma_{x}^{-1} & = & 
   \begin{pmatrix}
      0 & 1  \\ 
      1 & 0
   \end{pmatrix}   
      \begin{pmatrix}
     i \gamma I_{N/2} & U  \\ 
     U^{T} & - i \gamma I_{N/2}
   \end{pmatrix}
    \begin{pmatrix}
      0 & 1  \\ 
      1 & 0
   \end{pmatrix} 
   \\ & = &
   \begin{pmatrix}
      - i \gamma I_{N/2} & U^{T} \\ 
      U & i \gamma I_{N/2}
   \end{pmatrix}  =  H^{\dagger}.
\end{eqnarray*}

\begin{figure}
\begin{center} 
\includegraphics[width=0.35\textwidth]{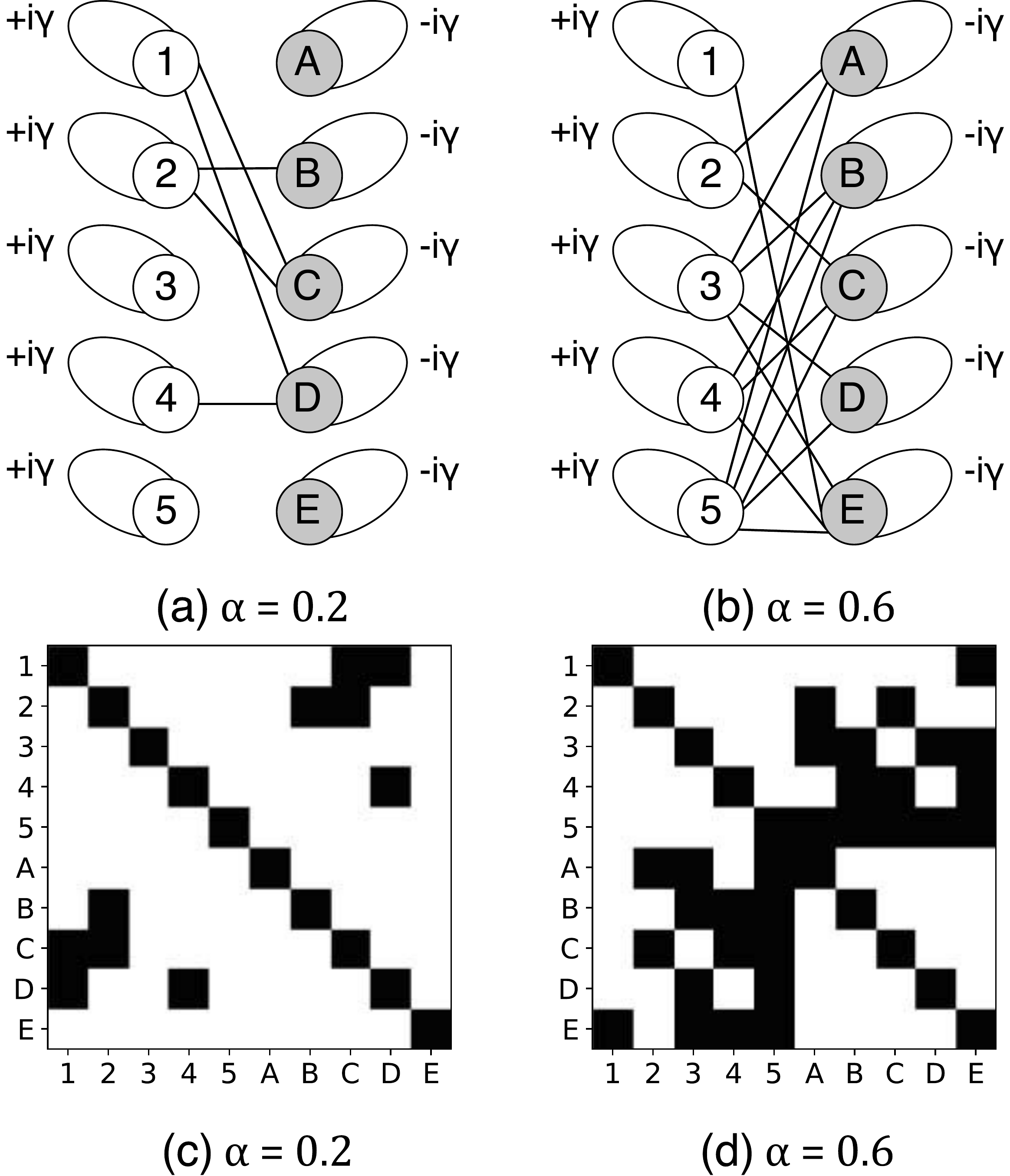}
\caption{(a-b) Examples of pseudo-Hermitian bipartite graphs. 
(c-d) Non-zero elements of the adjacency matrices corresponding to the graphs in the upper panels.
For comparison purposes here we choose the same parameter pairs $(N,\alpha)$ of the central panels
in Fig.~\ref{Fig01}.}
\label{Fig02}
\end{center}
\end{figure}

In Fig.~\ref{Fig02} we show some examples of pseudo-Hermitian 
bipartite tight-binding graphs together with their corresponding adjacency matrices. It is important to stress
that only in the trivial limit of $\alpha=0$ both the ${\cal PT}-$symmetric and pseudo-Hermitian 
bipartite graphs are equivalent.

Due to pseudo-Hermiticity, we expect to observe regions in the parameter space of our bipartite graph model where the 
spectrum of $H(\gamma,\alpha,N)$ is real~\citep{N19,M05} even when it is neither Hermitian nor ${\cal PT}-$symmetric.
We will numerically explore some spectral properties of $H(\gamma,\alpha,N)$ in Subsection~\ref{SS_PH}.

\subsection{The case $N=4$}

The random Hamiltonian of Eq.~(\ref{eq:1}) cannot be diagonalized analytically for large network sizes.
However, here we diagonalize $H(\gamma,\alpha,4)$ (where $N=4$ is the smallest non-trivial network size 
one can consider), so we can get some insight into the effect of the parameters $\gamma$ and $\alpha$
on the eigenvalues of $H(\gamma,\alpha,N)$.

Note that when $N=4$ the matrix $U$ is a $2 \times 2$ matrix. Thus, only four values of $\alpha$ are allowed:
0, 1/4, 1/2 and 1. Let us start by writing down the most general case, which corresponds to $H(\gamma,1,4)$ in 
the pseudo-Hermitian setup:
$$
   H^{pH}(\gamma,1,4) = 
   \begin{pmatrix}
      i \gamma & 0 & a & b  \\ 
         0 &  i \gamma & c & d \\
         a & c & -  i \gamma & 0\\
          b & d & 0 & -  i \gamma
   \end{pmatrix} 
$$
which has the eigenvalues
\begin{widetext}
\begin{eqnarray}
    \begin{aligned}
     \lambda_{1, 2}^{pH,\alpha=1}(\gamma) &= \pm \cfrac{1}{\sqrt{2}} \sqrt{a^2 +b^2+ c^2 + d^2 - \sqrt{ \left[ (b +c)^2 + (a-d)^2 \right] \left[ (b - c)^2 + (a+d)^2 \right] } - 2\gamma^2}, \\
    \lambda_{3, 4}^{pH,\alpha=1}(\gamma) &= \pm \cfrac{1}{\sqrt{2}} \sqrt{a^2 +b^2+ c^2 + d^2 + \sqrt{ \left[ (b +c)^2 + (a-d)^2 \right] \left[ (b - c)^2 + (a+d)^2 \right] } - 2\gamma^2} . 
    \end{aligned}
    \label{lambdapH}
\end{eqnarray}
Then, we set $b=c$ to get the ${\cal PT}-$symmetric setup:
\begin{eqnarray}
    \begin{aligned}
     \lambda_{1, 2}^{{\cal PT},\alpha=1}(\gamma) &= \pm \cfrac{1}{\sqrt{2}} \sqrt{a^2+2b^2+d^2 - (a+d)\sqrt{ 4b^2 + (a-d)^2 } - 2\gamma^2}, \\
    \lambda_{3, 4}^{{\cal PT},\alpha=1}(\gamma) &= \pm \cfrac{1}{\sqrt{2}} \sqrt{a^2+2b^2+d^2 + (a+d)\sqrt{ 4b^2 + (a-d)^2 } - 2\gamma^2} . 
    \end{aligned}
    \label{lambdapT}
\end{eqnarray}
\end{widetext}
Therefore, the eigenvalues for the cases $\alpha=0$, 1/4, 1/2 can be obtained from Eqs.~(\ref{lambdapH},\ref{lambdapT}) by
setting the values of $a$, $b$, $c$ and/or $d$ to zero. For example, in the trivial case of $\alpha=0$ we have
$
 \lambda_{1,2}^{pH,\alpha=0}(\gamma) =  \lambda_{3,4}^{pH,\alpha=0}(\gamma) =
 \lambda_{1,2}^{{\cal PT},\alpha=0}(\gamma) =  \lambda_{3,4}^{{\cal PT},\alpha=0}(\gamma) = \pm i\gamma
$.

From Eqs.~(\ref{lambdapH},\ref{lambdapT}) we can observe that for $\alpha>0$ the eigenvalues are either real 
pairs, positive and negative, or imaginary conjugates. That is, the eigenvalues fall on the real and imaginary axes 
for both setups. Also, for $\alpha=0$ the eigenvalues are double degenerate imaginary conjugates with magnitude
$\gamma$; that is $\gamma$ determines the imaginary spectral radius.

In what follows we numerically consider $N>4$.

\begin{figure*}
\begin{center} 
\includegraphics[width=0.7\textwidth]{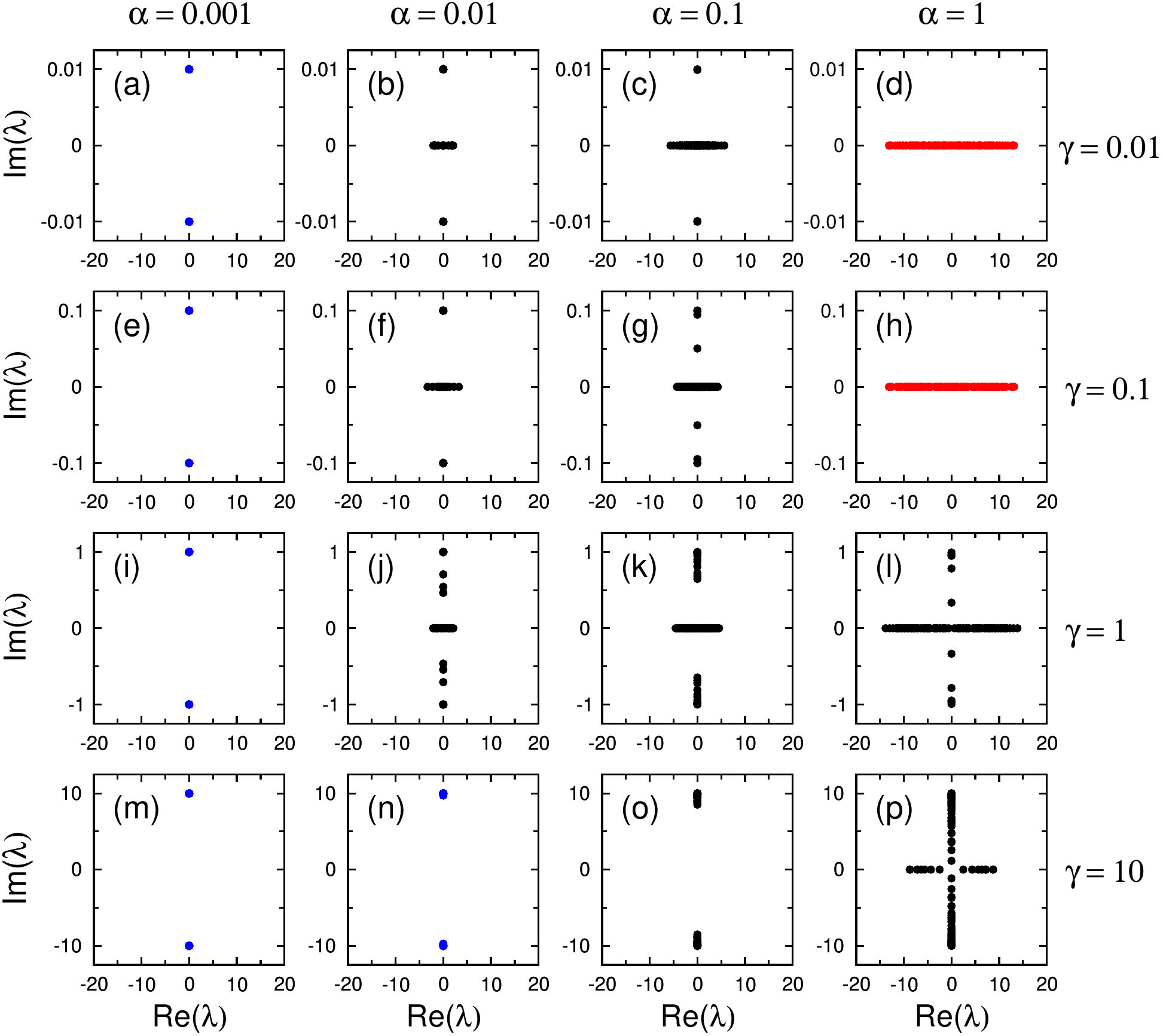}
\caption{Eigenvalues $\lambda$ of ${\cal PT}-$symmetric bipartite graphs of size $N=100$
and several combinations of $\alpha$ and $\gamma$. For easy identification we use blue (black) [red] when
the spectra is purely imaginary (complex) [purely real]. Single random graph realizations were used.}
\label{Fig03}
\end{center} 
\end{figure*}

\section{Spectral properties of $H(\gamma,\alpha,N)$}
\label{spectra}

Here we numerically diagonalize $H(\gamma,\alpha,N)$ in both ${\cal PT}-$symmetric
and pseudo-Hermitian setups to analyze, statistically, their spectra.
In particular, following~\cite{BG20}, we will compute the average fraction of real and imaginary eigenvalues, 
$\left< Q(\mbox{Re}) \right>$
and $\left< Q(\mbox{Im}) \right>$, respectively.
Here the average is taken over graph realizations; so, for every graph realization we compute
$$
Q(\mbox{Re}) = \frac{\mbox{number of real eigenvalues}}{N}
$$
and
$$
Q(\mbox{Im}) = \frac{\mbox{number of imaginary eigenvalues}}{N} .
$$
Evidently, $Q(\mbox{Re})+Q(\mbox{Im})=1$.
Also, we trivially expect $Q(\mbox{Re})\approx 0$ and $Q(\mbox{Im})\approx1$ for $\alpha\to 0$.
With respect to the various contexts described in the introduction, the stability of complex systems entails that the limit $\left< Q(\mbox{Re}) \right> \to 1$ is sought in the design (deterministic or evolutionary) of such systems.

\subsection{${\cal PT}-$symmetric Hamiltonian}
\label{SS_PT}

In Fig.~\ref{Fig03} we show the eigenvalues $\lambda$ in the complex plane of ${\cal PT}-$symmetric 
bipartite graphs of size $N=100$ for several combinations of $\alpha$ and $\gamma$.
From this figure we observe that:
(i) for $\alpha\to 0$ the spectrum is purely imaginary, see panels (a,e,i,m), moreover the spectrum
can also be purely imaginary for small $\alpha$ and large $\gamma$, see panel (n);
(ii) for $\alpha=1$ and $\gamma\to 0$ the spectrum is purely real, see panel (d), moreover the 
spectrum can also be purely real for large $\alpha$ and small $\gamma$, see panel (h);
(iii) for intermediate values of $\alpha$ and $\gamma$ the spectrum is mixed having real and imaginary
eigenvalues.
In either case the eigenvalues come in pairs, as expected due to particle-hole symmetry: if they are real there is one positive and one negative
with the same magnitude, if they are imaginary they are imaginary conjugates; so they fall on the
real and imaginary axes.
Even though we are reporting here the case $N=100$ only, we have observed the same panorama for
all the values of $N$ we have computed.

It is fair to recall that the panorama shown in Fig.~\ref{Fig03} has also been reported in~\cite{MHFO20} for the non-hermitian Su-Schrieffer-Heeger (SSH) model with random hopping terms, which is indeed ${\cal PT}-$symmetric. In fact, the SSH model is the particular case of our ${\cal PT}-$symmetric bipartite graph where the nodes form a linear chain with alternating loss and gain.

Note that each panel in Fig.~\ref{Fig03} reports a single random graph realization, this means that for 
the same combination of $(\alpha,\gamma)$ we may get different results; i.e.~a mixed spectra instead
of a purely real/imaginary spectra and vice-versa.
Therefore, in the following we focus on average properties.

\begin{figure*}
\begin{center} 
\includegraphics[width=0.65\textwidth]{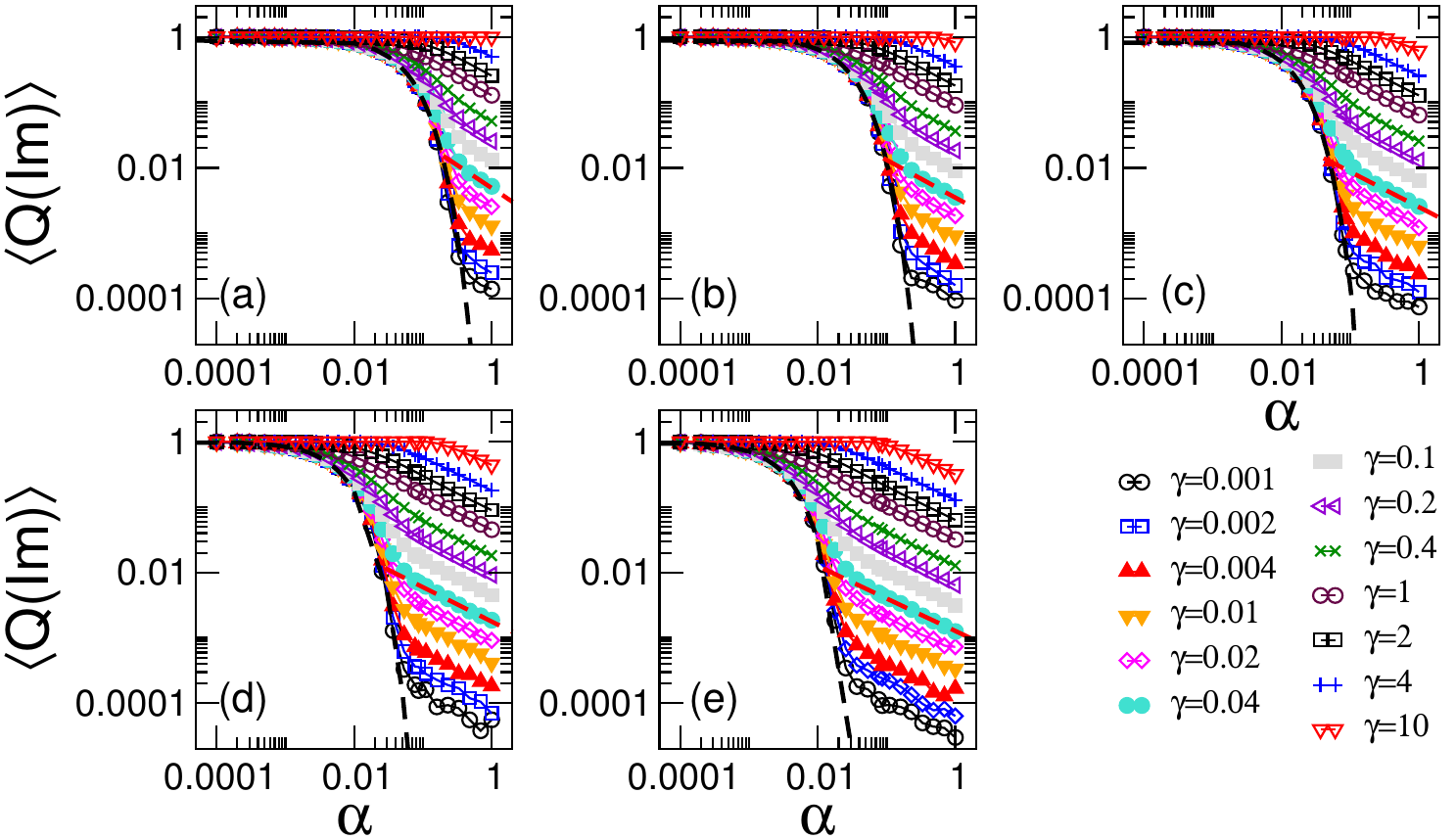}
\caption{Average fraction of imaginary eigenvalues $\left< Q(\mbox{Im}) \right>$ as a function of $\alpha$ for 
${\cal PT}-$symmetric bipartite graphs.
(a) $N=50$, (b) $N=100$, (c) $N=200$, (d) $N=400$, (e) $N=800$. 
Several values of $\gamma$ are shown. Black dashed lines are Eq.~(\ref{QImexpN}).
Red dashed lines are Eq.~(\ref{QImplNg}), particularly set at $\gamma=0.04$ as examples.
The averages are computed over $10^6/N$ random graphs.}
\label{Fig04}
\end{center} 
\end{figure*}
\begin{figure}
\begin{center} 
\includegraphics[width=0.45\textwidth]{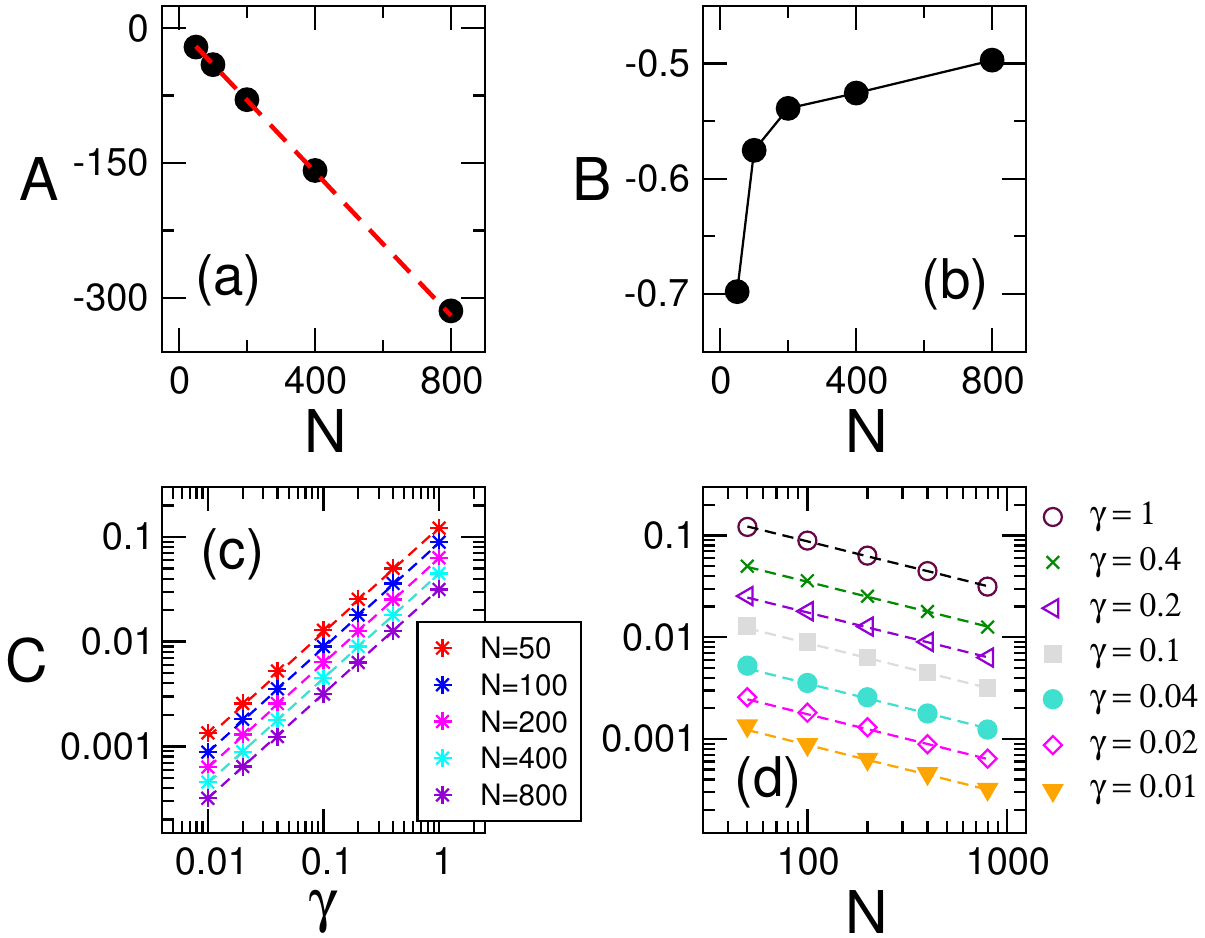}
\label{ajsh}
\caption{ (a) $\mathcal{A}$ as a function of $N$, (b) $\mathcal{B}$ as a function of $N$, 
(c) $\mathcal{C}$ as a function of $\gamma$, and (d) $\mathcal{C}$ as a function of $N$.
The dashed line in (a) is the best linear fitting giving ${\cal A}(N)= -1.2479-0.3919 N$.
The dashed lines in (c) and (d) correspond to power-law fittings indicating that 
$\mathcal{C}\propto \gamma$ and $\mathcal{C}\propto N^{-0.487}$, respectively.}
\label{Fig05}
\end{center} 
\end{figure}

Now in Fig.~\ref{Fig04} we plot the average fraction of imaginary eigenvalues 
$\left< Q(\mbox{Im}) \right>$ as a function of $\alpha$ for ${\cal PT}-$symmetric bipartite graphs.
Each panel reports several values of $\gamma$ for a fixed graph size $N$.
As anticipated in Fig.~\ref{Fig03}, here we observe that $\left< Q(\mbox{Im}) \right>$ is a monotonically decreasing 
function of $\alpha$, with $\left< Q(\mbox{Im}) \right>\approx 1$ for $\alpha\to 0$ and $\left< Q(\mbox{Im}) \right>\to 0$ 
for $\alpha\to 1$ when $\gamma$ is small enough. 
Moreover, the curves $\left< Q(\mbox{Im}) \right>$ vs.~$\alpha$ show two regimes (better
appreciated for small $\gamma$):
an initial exponential decrease, 
\begin{equation}
\left< Q(\mbox{Im}) \right> = \exp(-{\cal A} \alpha) \, ,
\label{QImexp}
\end{equation}
followed by a power-law decrease,
\begin{equation}
\left< Q(\mbox{Im}) \right> = {\cal C} \alpha^{-\cal B} \, .
\label{QImpl}
\end{equation}

By fitting the exponential decay of the curves $\left< Q(\mbox{Im}) \right>$ vs.~$\alpha$ with 
Eq.~(\ref{QImexp}) for $\gamma=0.001$, we found that ${\cal A}$ depends on the graph size. 
Moreover, ${\cal A}(N)\approx -1.2479-0.3919 N\approx -0.4N$; see Fig.~\ref{Fig05}(a). Indeed, 
the black dashed lines in all panels of Fig.~\ref{Fig04} are
\begin{equation}
\left< Q(\mbox{Im}) \right> \approx \exp(-0.4 N \alpha) \, ,
\label{QImexpN}
\end{equation}
which describe reasonably well the initial exponential decay of $\left< Q(\mbox{Im}) \right>$.
Then, we fitted the power-law decay of the curves $\left< Q(\mbox{Im}) \right>$ vs.~$\alpha$
with Eq.~(\ref{QImpl}). We found that $\mathcal{B}$ does not depend 
on $\gamma$ but depends on $N$ in a highly non-trivial way, see Fig.~\ref{Fig05}(b), while
$\mathcal{C}$ depends on both $N$ and $\gamma$, see Figs.~\ref{Fig05}(c,d).
Therefore, we can write
\begin{equation}
\left< Q(\mbox{Im}) \right> \approx 0.825 \gamma N^{-0.487} \alpha^{-\cal B} \, ;
\label{QImplNg}
\end{equation}
which provides a good description of the power-law decay regime of $\left< Q(\mbox{Im}) \right>$,
as the red dashed lines in Fig.~\ref{Fig04} show.

\begin{figure*}
\begin{center} 
\includegraphics[width=0.65\textwidth]{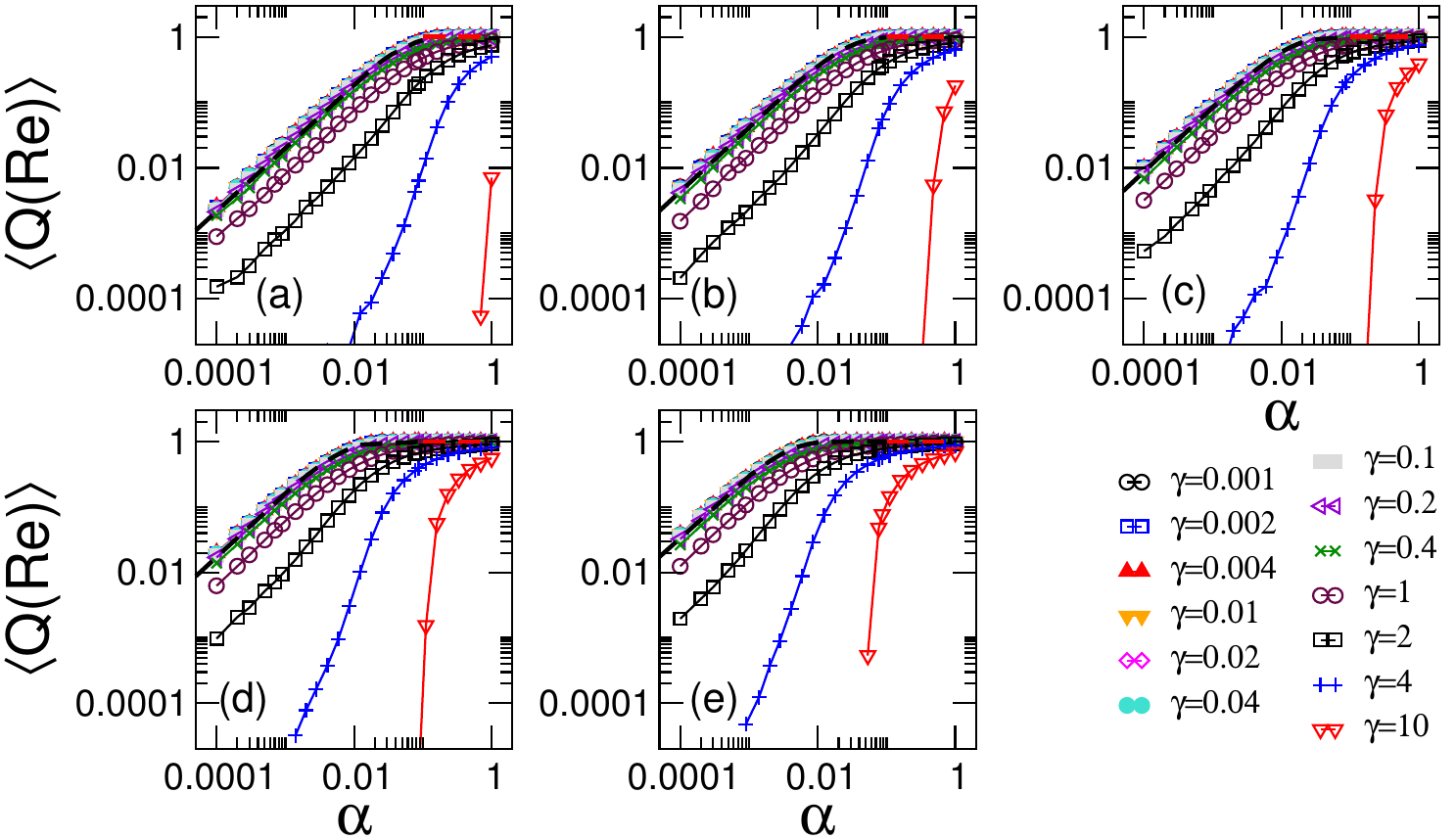}
\caption{Average fraction of real eigenvalues $\left< Q(\mbox{Re}) \right>$ as a function of $\alpha$ of 
${\cal PT}-$symmetric bipartite graphs.
(a) $N=50$, (b) $N=100$, (c) $N=200$, (d) $N=400$, (e) $N=800$. 
Several values of $\gamma$ are shown. Black dashed lines are Eq.~(\ref{QReexpN}).
Red dashed lines are Eq.~(\ref{QReplNg}).
The averages are computed over $10^6/N$ random graphs.}
\label{Fig06}
\end{center} 
\end{figure*}

In addition, in Fig.~\ref{Fig06} we plot the average fraction of real eigenvalues 
$\left< Q(\mbox{Re}) \right>$ as a function of $\alpha$ of ${\cal PT}-$symmetric bipartite graphs.
In fact, since $\left< Q(\mbox{Re}) \right>$ and $\left< Q(\mbox{Im}) \right>$ are complementary quantities,
i.e.~$\left< Q(\mbox{Re}) \right>+\left< Q(\mbox{Im}) \right>=1$, $\left< Q(\mbox{Re}) \right>$ is a monotonically
increasing function of $\alpha$, with $\left< Q(\mbox{Re}) \right>\approx 0$ for $\alpha\to 0$ and 
$\left< Q(\mbox{Re}) \right>\to 1$ for $\alpha\to 1$ when $\gamma$ is small enough.
Moreover, from Eqs.~(\ref{QImexpN}) and~(\ref{QImplNg}) we can write
\begin{equation}
\left< Q(\mbox{Re}) \right> \approx 1-\exp(-0.4 N \alpha) \, ,
\label{QReexpN}
\end{equation}
and
\begin{equation}
\left< Q(\mbox{Re}) \right> \approx 1-0.825 \gamma N^{-0.487} \alpha^{-\cal B} \, ,
\label{QReplNg}
\end{equation}
which describes well the curves $\left< Q(\mbox{Re}) \right>$ vs.~$\alpha$ for small $\gamma$;
see the black and red dashed lines in Fig.~\ref{Fig06}, respectively.

\begin{figure}
\begin{center} 
\includegraphics[width=0.48\textwidth]{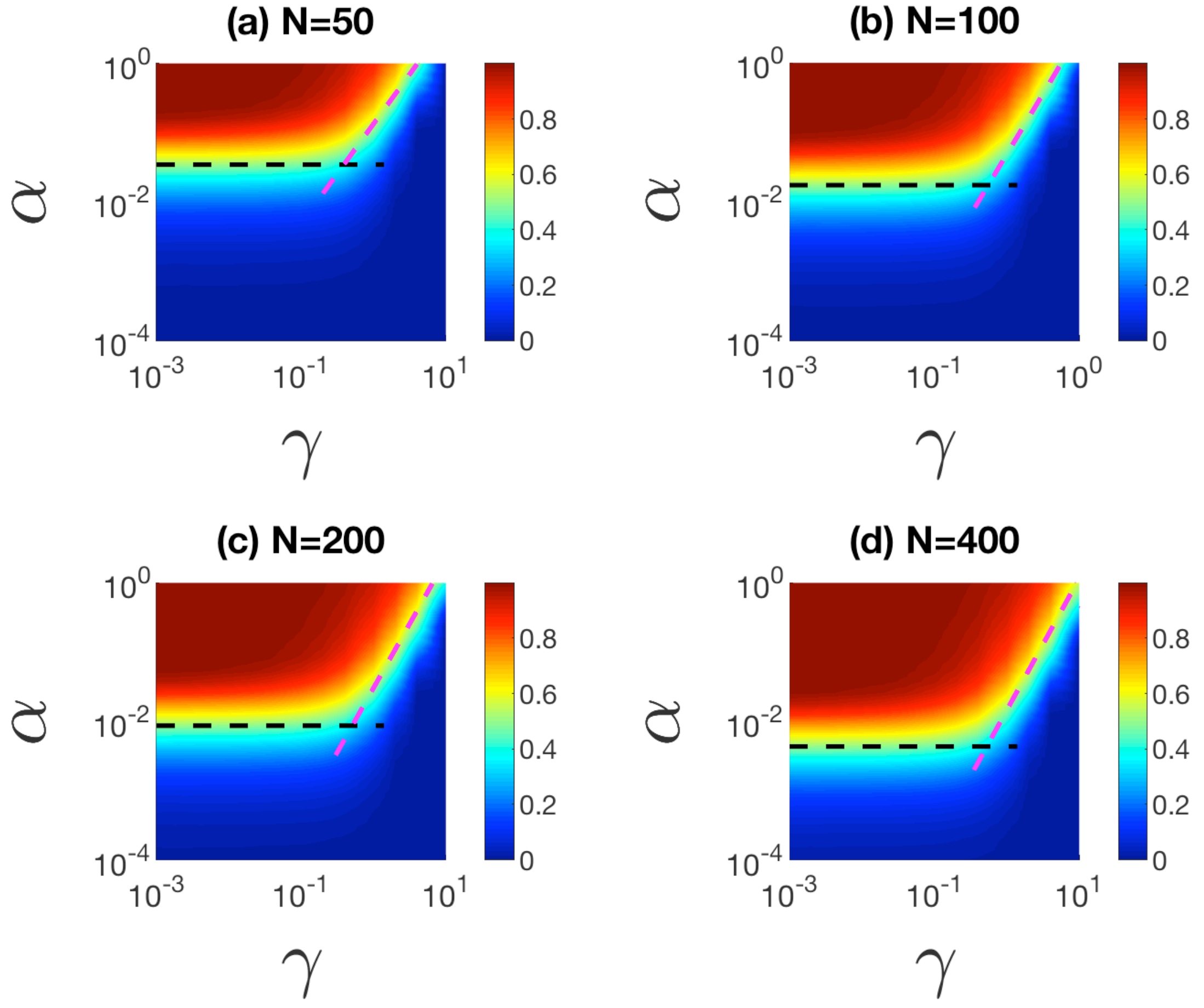}
\caption{Average fraction of real eigenvalues $\left< Q(\mbox{Re}) \right>$ in the $\gamma\alpha-$plane 
of ${\cal PT}-$symmetric bipartite graphs of different sizes $N$.
Black and red dashed lines are Eqs.~(\ref{alphac}) and~(\ref{alphacgammac}), respectively and devide, 
approximately, the $\gamma\alpha-$plane in two regions: one where the spectra has more real than 
imaginary eigenvalues (above the dashed lines) and the other where the imaginary eigenvalues 
dominate over the real ones (below the dashed lines).}
\label{Fig07}
\end{center} 
\end{figure}

Finally, in Fig.~\ref{Fig07} we present again $\left< Q(\mbox{Re}) \right>$ but in the $\gamma\alpha-$plane of 
${\cal PT}-$symmetric bipartite graphs of different sizes. Here dark red means a real spectra (upper left
corner of the colormaps) while dark blue represents imaginary spectra (lower part of the colormaps).
Indeed, the green stripe in these colormaps, which marks 
$\left< Q(\mbox{Re}) \right> \approx \left< Q(\mbox{Im}) \right> \approx 1/2$, divides
the $\gamma\alpha-$plane in two well defined regions: one where the spectra has more real than imaginary 
eigenvalues (above the green stripe) and the other where the imaginary eigenvalues dominate over the 
real ones (below the green stripe).
Moreover, we can estimate the location of the green stripe in the $\gamma\alpha-$plane, which is parametrized
by $\gamma_c$ and $\alpha_c$, as follows.
First, we notice that for small $\gamma$ the green stripe is characterized by a constant value of $\alpha_c$.
This value of $\alpha_c$ can be estimated by equating $\left< Q(\mbox{Im}) \right> = 1/2$ or 
$\left< Q(\mbox{Re}) \right>=1/2$ in~(\ref{QImexpN}) or~(\ref{QReexpN}), respectively. We then obtain  
\begin{equation}
\alpha_c \approx - \frac{1}{0.4 N} \ln \frac{1}{2} \approx 1.73 N^{-1} \, .
\label{alphac}
\end{equation} 
Indeed, Eq.~(\ref{alphac}) is shown as the black dashed lines on the colormaps of Fig.~\ref{Fig07}.
Second, we observe that for large values of $\gamma$ the green stripe depends on both $\alpha$ and $\gamma$.
These values of $\alpha_c$ and $\gamma_c$ characterizing the green stripe can be estimated by equating 
$\left< Q(\mbox{Im}) \right> = 1/2$ or $\left< Q(\mbox{Re}) \right>=1/2$ in~(\ref{QImplNg}) or~(\ref{QReplNg}), 
respectively. We then get  
\begin{equation}
\alpha_c \gamma_c^{1/{\cal B}} \approx (0.6 N^{0.487})^{1/{\cal B}} \, .
\label{alphacgammac}
\end{equation} 
Equation~(\ref{alphacgammac}) is shown as the red dashed lines on the colormaps of Fig.~\ref{Fig07}.
It is clear from this figure that the region where the real eigenvalues dominate over the imaginary ones
grows with $N$.

\subsection{Pseudo-Hermitian Hamiltonian}
\label{SS_PH}

For the pseudo-Hermitian setup we observe a qualitative similar panorama to that described in the previous
Subsection for the ${\cal PT}-$symmetric setup, with slight quantitative differences.
Namely:
\begin{itemize}
\item[(i)] for $\alpha\to 0$ the spectrum is purely imaginary;
\item[(ii)] for $\alpha=1$ and $\gamma\to 0$ the spectrum is purely real;
\item[(iii)] for intermediate values of $\alpha$ and $\gamma$ the spectrum is mixed having real and 
imaginary eigenvalues;
\item[(iv)] in either case the eigenvalues fall on the real and/or the imaginary axes;
\item[(v)] $\left< Q(\mbox{Im}) \right>$ is a monotonically decreasing 
function of $\alpha$, with $\left< Q(\mbox{Im}) \right>\approx 1$ for $\alpha\to 0$ and 
$\left< Q(\mbox{Im}) \right>\to 0$ for $\alpha\to 1$ when $\gamma$ is small enough;
\item[(vi)] the curves $\left< Q(\mbox{Im}) \right>$ vs.~$\alpha$ show an initial exponential decrease
\begin{equation}
\left< Q(\mbox{Im}) \right> \approx \exp(-0.4 N \alpha) \, ,
\label{QImexpNpH}
\end{equation}
followed by the power-law
\begin{equation}
\left< Q(\mbox{Im}) \right> \approx 1.12 \gamma^{0.9} N^{-0.54} \alpha^{-\cal B} \, ,
\label{QImplNgpH}
\end{equation} 
where ${\cal B}$ is a highly non-trivial function of $N$ (very similar to Fig.~\ref{Fig05}(b));
\item[(vii)] $\left< Q(\mbox{Re}) \right>$ is a monotonically
increasing function of $\alpha$, with $\left< Q(\mbox{Re}) \right>\approx 0$ for $\alpha\to 0$ and 
$\left< Q(\mbox{Re}) \right>\to 1$ for $\alpha\to 1$ when $\gamma$ is small enough.
\end{itemize}

\begin{figure}
\begin{center} 
\includegraphics[width=0.48\textwidth]{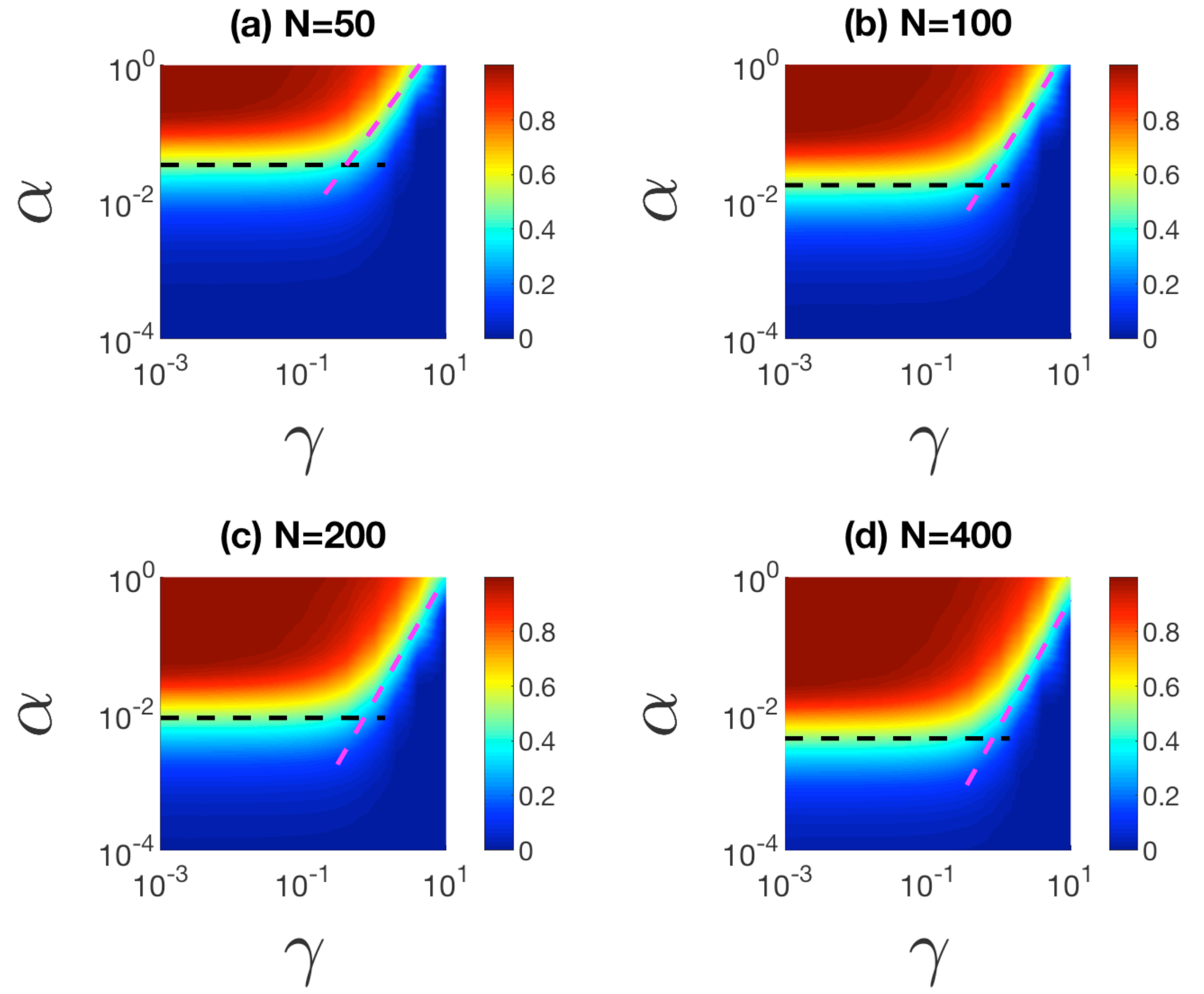}
\caption{Average fraction of real eigenvalues $\left< Q(\mbox{Re}) \right>$ in the $\gamma\alpha-$plane 
of pseudo-Hermitian bipartite graphs of different sizes $N$.
Black and red dashed lines are Eqs.~(\ref{alphacpH}) and~(\ref{alphacgammacpH}), respectively.}
\label{Fig08}
\end{center} 
\end{figure}

Then, for comparison purposes, in Fig.~\ref{Fig08} we present $\left< Q(\mbox{Re}) \right>$ in the 
$\gamma\alpha-$plane of pseudo-Hermitian bipartite graphs of different sizes. 
There, black and red dashed lines are, respectively,
\begin{equation}
\alpha_c \approx 1.73 N^{-1}
\label{alphacpH}
\end{equation} 
and
\begin{equation}
\alpha_c \gamma_c^{0.9/{\cal B}} \approx (0.45 N^{0.54})^{1/{\cal B}} \, ,
\label{alphacgammacpH}
\end{equation} 
which divide the $\gamma\alpha-$plane in two regions: one where the spectra has more 
real than imaginary eigenvalues (above the dashed lines) and the other where the imaginary eigenvalues 
dominate over the real ones (below the dashed lines).
As for ${\cal PT}-$symmetric bipartite graphs, here in the pseudo-Hermitian setup, the region in the $\gamma\alpha-$plane where  real eigenvalues dominate over imaginary ones grows with the graph size.

\section{Discussion and conclusions}
\label{conclusions}

We have studied bipartite tight-binding random graphs with balanced on-site loss and gain.
The graph is formed by two sets of equal size: one set containing the nodes with loss, the other set having 
nodes with gain; the nodes of different sets are connected randomly while connections among nodes of the
same set are forbidden.
Thus, our graph Hamiltonian depends on three parameters: the loss/gain strength $\gamma$, the connectivity 
between the two sets $\alpha$, and the graph size $N$.

By analyzing the symmetries of $H(\gamma,\alpha,N)$, we recognized pseudo-Hermiticity (for general 
undirected-graph setups) and ${\cal PT}-$symmetry (for given graph setups) which may produce real
eigenvalues even when $H(\gamma,\alpha,N)$ is non-hermitian. Then we numerically focused on the 
average fractions of imaginary and real eigenvalues of $H(\gamma,\alpha,N)$ ($\left< Q(\mbox{Im}) \right>$ 
and $\left< Q(\mbox{Re}) \right>$, respectively, with $\left< Q(\mbox{Im}) \right>+\left< Q(\mbox{Re}) \right>=1$) 
as a function of the parameter set $\{\gamma,\alpha,N\}$. In both setups, pseudo-Hermiticity and 
${\cal PT}-$symmetric, we numerically found that 
$\left< Q(\mbox{Im}) \right>$ [resp. $\left< Q(\mbox{Re}) \right>$] is a monotonically decreasing [resp. increasing]
function of $\alpha$, with $\left< Q(\mbox{Im}) \right>\approx 1$ [resp. $\left< Q(\mbox{Re}) \right>\approx 0$] 
for $\alpha\to 0$ and $\left< Q(\mbox{Im}) \right>\to 0$ [resp. $\left< Q(\mbox{Re}) \right>\to 1$] for $\alpha\to 1$ 
when $\gamma$ is small enough. Moreover, we observed that $\left< Q(\mbox{Im}) \right>$ decreases
with $\alpha$ in two clear forms: first exponentially and in a later regime as a power-law.

\begin{figure}
\begin{center} 
\includegraphics[width=0.5\textwidth]{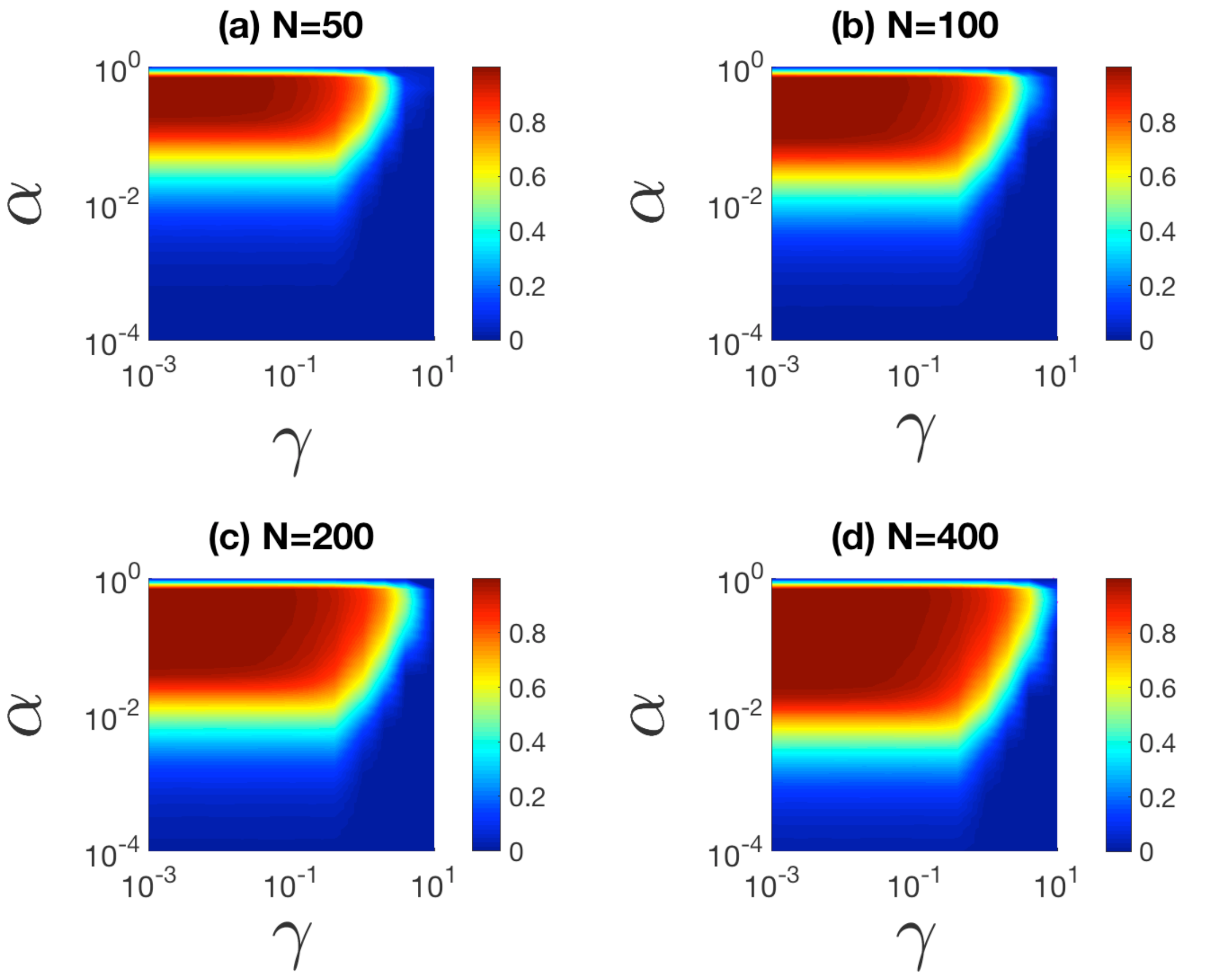}
\caption{Average fraction of real eigenvalues $\left< Q(\mbox{Re}) \right>$ in the $\gamma\alpha-$plane of 
${\cal PT}-$symmetric bipartite binary graphs of different sizes.}
\label{Fig09}
\end{center} 
\end{figure}

Another scenario of interest may be the binary setup, where the hopping integrals between sites $n$ and $m$
are all set to one: $u_{mn}=u_{nm}=1$ in Eq.~(\ref{TBH}). This produces binary $U$ matrices; 
i.e.~$H_\alpha(\alpha,N)$ in Eq.~(\ref{eq:1}) becomes a standard graph adjacency matrix.
We have observed a very similar panorama to that discussed above, when the hopping integrals are random
variables, except in the limit $\alpha\to 1$.
For example, for $N=4$ we have
\begin{eqnarray}
    \begin{aligned}
     \lambda_{1, 2}^{pH,\alpha=1}(\gamma) &= \lambda_{1, 2}^{{\cal PT},\alpha=1}(\gamma) = \pm i\gamma, \\
    \lambda_{3, 4}^{pH,\alpha=1}(\gamma) &= \lambda_{3, 4}^{{\cal PT},\alpha=1}(\gamma) = \pm \sqrt{4 - \gamma^2} ,
    \end{aligned}
    \label{lambdapHb}
\end{eqnarray}
which are obtained from Eqs.~(\ref{lambdapH},\ref{lambdapT}) with $a=b=c=d=1$. From~(\ref{lambdapHb}) it is clear that real eigenvalues are forbidden when $\alpha=1$ in the $N=4$ case.
Moreover, we have also numerically explored the binary case in both setups, pseudo-Hermiticity and ${\cal PT}-$symmetric, for large $N$ (noting in passing the direct relation of the limit of such binary cases with various flavours of Hamiltonians that deal with ``bipartite crossing manifolds'', which have been studied first in quantum mechanics but also more recently in photonics~\cite{Ostrovsky1995,Ostrovsky1995b,BenistyPRA2011,KurtJOSA2008,KhayamPNFA2010}).
In Fig.~\ref{Fig09} we show the average fraction of real eigenvalues 
$\left< Q(\mbox{Re}) \right>$ in the $\gamma\alpha-$plane of ${\cal PT}-$symmetric bipartite binary graphs of 
different sizes (equivalent figures are obtained for pseudo-Hermitian binary graphs). Note that the colormaps of this figure are similar to those in Figs.~\ref{Fig07} and~\ref{Fig08},
except at $\alpha\to 1$ where the dark blue stripe indicates the absence of real eigenvalues.

Concerning the relevance of these results to current studies, they could be guided by the following considerations: (i) The dynamics should be very different in regimes dominated by mostly-imaginary vs.~mostly-real spectra, so a prerequisite is to assess how such broadly different behaviours are manifested in realistic systems; (ii) We have not investigated eigenvectors yet. They would of course be very important as they will sample eigenvalues. For instance, balancing energy fluxes in an open system is certainly an interesting working point, that selects eigenvectors and eigenvalues combinations and might screen the existence of extreme ones; (iii) The possibility to play with realistic systems having large $N$ is key to check some of the scaling laws and of the fraction of eigenvalues of a given kind. 
Photonics-based realizations with randomness, such as ``random lasers on graphs'' \cite{Gaio2019}, look like very good candidates in this respect. In biology, units with a well-defined role are not obviously the proteins, that are liable to a number of partial interactions. Rather, the organites that manage the energy balance of a cell or of an organ (mitochondria, chloroplasts, etc.) could show more clear-cut characteristics. As for the applicability to economy, there are probably sectors where competition is more clear-cut and others where cooperation is pervasive. However, to have a large number of small units requires to deal with less formal ones. Tracking some specific goods that must obviously relate to energy, such as oil, or electricity, could help clarifying the tangled web of generic economical/social interaction into a more binary and bipartite pattern, especially in areas with many small producers, such as solar energy production from roofs.

A different direction would be to investigate the susceptibility of the results to perturbations. For instance, is the system more prone to evolve when almost all its eigenvalues are of the same kind (because of the likeliness of generating an extra minority eigenvalue, thus sizably influencing the small fraction)? Or is it more prone to evolve mid-way along the real/imaginary transition (see Fig.~\ref{Fig03})?

Whatever the answers, the tools presented here constitute a useful guidance in the area of random bipartite non-hermitian systems.

\begin{acknowledgments}
J.A.M.-B. thanks support from CONACyT (Grant No. 286633), CONACyT-Fronteras (Grant No. 425854), 
VIEP-BUAP (Grant No. 100405811-VIEP2022), and Laboratorio Nacional de Superc\'omputo del Sureste 
de M\'exico (Grant No. 202201007C), Mexico.
\end{acknowledgments}

\end{document}